\documentclass[twocolumn,showpacs,showkeys,amsmath,amssymb,aps,prb]{revtex4-1}
\usepackage{graphicx}
\newcommand{\lk}{\left(}
\newcommand{\rk}{\right)}
\newcommand{\lab}{\left|}
\newcommand{\rab}{\right|}
\newcommand{\lkv}{\left[}
\newcommand{\rkv}{\right]}
\newcommand{\lfi}{\left\{}
\newcommand{\rfi}{\right\}}
\newcommand{\be}{\begin{equation}}
\newcommand{\ee}{\end{equation}}
\newcommand{\vh}{\mathbf{h}}
\newcommand{\vz}{\mathbf{z}}
\newcommand{\vS}{\mathbf{S}}

\begin{document}
\title{Exact ground state of the Shastry-Sutherland lattice with classical Heisenberg spins}
\author{Alexei Grechnev}
\email{shrike4625@yahoo.com}
\affiliation{B. Verkin Institute for Low Temperature Physics and
Engineering of the National Academy of Sciences of Ukraine, 47 Lenin Avenue, Kharkiv 61103, Ukraine}

\begin{abstract}
An exact analytical solution of the ground state problem of the isotropic classical
Heisenberg model on the Shastry-Sutherland lattice in external magnetic field $H$
is found for arbitrary ratio of diagonal and edge exchange constants $J_2/J_1$.
The phase diagram of this model in the ($J_2/J_1, H/J_1$) plane
is presented. It includes spin-flop, spin-flip and umbrella phases.
The magnetization curves are found to be linear until saturation. 
It is shown numerically that the inclusion of the easy-axis anisotropy
into the model leads to the appearance of the $1/3$ magnetization plateau,
corresponding to the collinear up-up-down spin structure. This explains
the appearance of the $1/3$ magnetization plateau in rare earth tetraborides
RB$_4$. In particular, magnetization curve of the compound HoB$_4$ is explained.
\end{abstract}


\maketitle

\section{Introduction}

\begin{figure}
\includegraphics[scale=0.45]{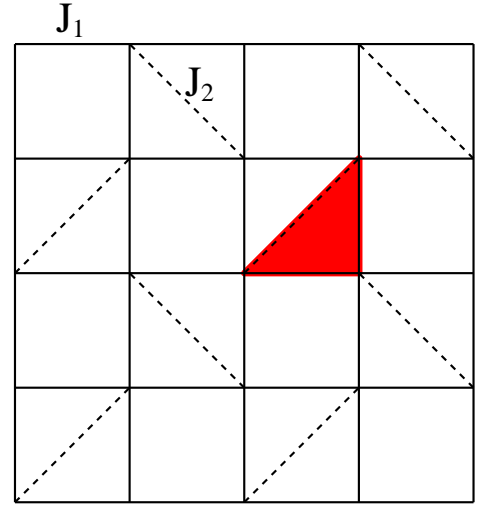}
\caption{\label{f:ssl} (Color online) Shastry-Sutherland lattice. The red (gray) triangle 
marks the elementary cluster of the SSL.}
\end{figure}

Shastry-Sutherland lattice (SSL) was first introduced in 
the work of Shastry and Sutherland \cite{shastry81physicabc108:1069} as a purely
theoretical example of a two-dimensional frustrated spin system. SSL is a square lattice with classical 
Heisenberg, quantum Heisenberg, or Ising spin $\vS_i$
at every lattice site $i$, with antiferromagnetic
(AFM) exchange $J_1$ along the edges, and AFM exchange $J_2$ along certain diagonals,
as shown in Fig. \ref{f:ssl}. Its Hamiltonian (in the presence of the external field $H$ 
directed along the $z$-axis) is
\be
\label{hamilt}
\mathcal{H} = J_1 \sum_{edges} \vS_i \cdot \vS_j + J_2 \sum_{diagonal} \vS_i \cdot \vS_j  -
H \sum_i S_i^z,
\ee
and the Hamiltonian normalized by $J_1$ is
\be
\label{hamilt2}
\widetilde{\mathcal{H}} \equiv \frac {\mathcal{H}}{J_1} = \sum_{edges} \vS_i  \cdot \vS_j + 
\rho \sum_{diagonal} \vS_i  \cdot  \vS_j  - h \sum_i S_i^z,
\ee
where we have introduced the notations
\be
\rho \equiv \frac {J_2}{J_1}, 
\quad h \equiv \frac H{J_1}.
\ee

Surprisingly, ten years after the work of Shastry and Sutherland,
SSL has been experimentally realized in SrCu$_2$(BO$_3$)$_2$ \cite{smith91jssc93:430},
which has a layered structure, with each Cu$^{2+}$ ion carrying 
spin $S=1/2$. If exchange interactions with only two spheres of nearest neighbors are included,
the magnetic lattice of SrCu$_2$(BO$_3$)$_2$ is equivalent to SSL. At low temperatures SrCu$_2$(BO$_3$)$_2$ exhibits
a sequence of magnetization plateaus at fractional values of the saturation magnetization $M_s$
\cite{kageyama99prl82:3168,onizuka00jpsj69:1016,kodama02science298:395}. A number of theories
of this phenomenon has been proposed \cite{miyahara00prb61:3417,suzuki09prb80:180405R,dorier08prl101:250402}.

Similar fractional magnetization plateaus have been observed recently in rare earth tetraborides 
\cite{michimura06physicab378:596,yoshii08prl101:087202,siemensmeyer08prl101:177201,matas10jpcs200:032041}
RB$_4$, where R=Tm, Er or Ho,
where the rare earth ions also form layered structure equivalent to SSL.
The important difference is that while Cu$^{2+}$ ions in SrCu$_2$(BO$_3$)$_2$ have spins $s=1/2$, the 
rare earth ions in RB$_4$ systems have large spins,
which can be treated as classical ones. These compounds 
also possess a strong easy-axis magnetocrystalline anisotropy.

\begin{figure}
\includegraphics[scale=0.35]{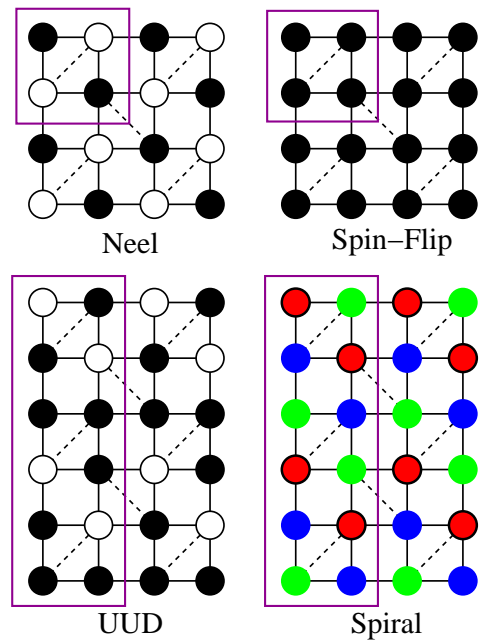}
\caption{\label{f:structures} (Color online) Neel, spin-flip, UUD and spiral (for $J_2/J_1=2$) spin structures. 
Black and white circles denote spins pointing up and down respectively. Red, green and blue circles
denote three different directions of the spiral structure, with angle $120^{\circ}$ between each two 
directions. The magnetic unit cells are shown with purple (gray) rectangles.}
\end{figure}

The discovery of magnetization plateaus in RB$_4$ compounds led to a 
number of theoretical and computational studies of SSL with classical
Heisenberg \cite{moliner09prb79:144401,qin11jap109:07E103,slavin11fnt37:1264,huo12arxiv:1211.3872} and
Ising \cite{huang12jpcm24:386003,chang09prb79:104411,meng08prb78:224416,
dublenych12prl109:167202,liu09arxiv:0904.3018,farkasovsky10prb82:054409} spins.
Deep understanding of these two models is vital for explaining 
the peculiar magnetization curves of RB$_4$ systems, as they can serve
as the foundation stones on which more complicated models with additional interactions
can be built. 
A major breakthrough for Ising SSL in external magnetic field came very recently as
its exact ground state has been found analytically \cite{dublenych12prl109:167202}.
This model gives a single $M/M_s=1/3$ magnetization plateau, which corresponds
to the so-called up-up-down (UUD) phase (Fig. \ref{f:structures}, lower left).
On the other hand, a
Monte Carlo simulation for the classical Heisenberg SSL \cite{moliner09prb79:144401}
found magnetization curves with no steps. However, when the easy-axis
anisotropy was included in the model \cite{qin11jap109:07E103,slavin11fnt37:1264,huo12arxiv:1211.3872},
the $1/3$ UUD plateau appeared for a certain range of $J_2/J_1$. 
Magnetization steps other than $1/3$ do not appear for either Ising or Heisenberg SSL. 
Additional exchange or dipolar interactions
\cite{huang12jpcm24:386003,dublenych12prl109:167202,farkasovsky10prb82:054409,huo12arxiv:1211.3872}
or lattice disorder \cite{qin11jap109:07E103} were employed to account
for those plateaus, in particular the large $1/2$ plateau found in TmB$_4$ (Ref. \onlinecite{huo12arxiv:1211.3872}).
If these additional interactions are included,
Ising model on SSL essentially succeeds in explaining
the appearance of fractional magnetization plateaus in RB$_4$. 
A typical zero-temperature magnetization curve $M(H)$ of the Ising SSL has the "staircase" shape,
namely it consists of horizontal magnetization
steps (including the $M=0$ and $M=M_s$ ones) separated by first-order
phase transitions (vertical segments of the $M(H)$ curve). This is very similar
to the experimental $M(H)$ curve of TmB$_4$ (Ref. \onlinecite{siemensmeyer08prl101:177201}).
Anisotropic Heisenberg SSL \cite{qin11jap109:07E103,slavin11fnt37:1264,huo12arxiv:1211.3872}, on the other hand,
gives smoother $M(H)$ curves with inclined regions, which were experimentally
observed for HoB$_4$ (Ref. \onlinecite{matas10jpcs200:032041}) and possibly
ErB$_4$ (Ref. \onlinecite{michimura06physicab378:596}). Of course, the latter model
can be applied to TmB$_4$ as well \cite{huo12arxiv:1211.3872}, provided that the
anistropy constant is lange enough.

While the exact ground state of the Ising SSL in magnetic field has been found,
the complete understanding of the classical Heisenberg SSL is still lacking.
In particular, its exact ground state has not been determined, except for special 
cases $H=0$ (Ref. \onlinecite{shastry81physicabc108:1069}) and $J_2/J_1=2$ (Ref. \onlinecite{moliner09prb79:144401}).
The Monte Carlo simulations \cite{moliner09prb79:144401,qin11jap109:07E103,slavin11fnt37:1264,huo12arxiv:1211.3872}
mainly focused on the special point $J_2/J_1=2$ or its vicinity, while
the phase diagram in the ($J_2/J_1,H/J_1$) plane has never been published.
The present paper is an attempt to clarify these issues.
Its goal is to study in detail the ground-state
problem of the isotropic classical Heisenberg SSL
and to determine the phase diagram of this model
in the ($\rho,h$) plane. The very interesting problem of the 
classical Heisenberg SSL with easy-axis anisotropy is also
addressed briefly in the present paper. 

The paper is organized as follows. Section \ref{sec:phase} introduces 
different possible magnetic structures of the SSL and presents its phase diagram in the ($\rho,h$) plane. 
In section \ref{sec:sol} the
exact ground state of the SSL is calculated and it is proven that the three phases of the previous
section are indeed the ground state spin structures. 
Section \ref{sec:num} checks the exact result with numerical 
simulation and examines the effect of the uniaxial anisotropy
on the magnetization curves. The experimental magnetization curve of
HoB$_4$ is also analyzed in this section. It is followed by a conclusion.

\section{\label{sec:phase}Magnetic structures and the phase diagram}
In this and the following sections we consider the problem of the ground state of the SSL
as the function of two parameters: $h \equiv H/J_1$ and $\rho \equiv J_2/J_1$.
In the absence of external magnetic field ($h=0$) the
problem has been solved in Ref. \onlinecite{shastry81physicabc108:1069}. For $\rho<1$
($J_2 < J_1$) the ground state
is the regular Neel AFM state (Fig. \ref{f:structures}, upper left) with the energy per lattice site
\be
\label{e-neel}
\epsilon_{Neel}=-2 + \frac {\rho}2.
\ee
The Neel state satisfies all $J_1$ exchange interactions, but not $J_2$ ones.
For $\rho>1$ ($J_2 > J_1$) the ground state is the so-called spiral state.
The angle between neighboring spins is $\pi-\triangle\phi$ along the edges, and
 $2 \triangle\phi$ along the diagonals, where
 $\triangle\phi=\cos^{-1} (1/\rho)=\cos^{-1} (J_1/J_2) $. The energy of this structure is
\be
\epsilon_{Spiral}=-\frac{1}{\rho} - \frac {\rho}2.
\ee
Such configuration can be constructed in different ways, leading to a
degeneracy \cite{shastry81physicabc108:1069,moliner09prb79:144401}.
In general, it is incommensurate with the crystal lattice, 
but for chosen values of $\rho$, namely for $\cos^{-1} (1/\rho) = \pi m/n$ with integer $m,n$,
periodic spirals can be realized. 
One possible spiral configuration for $\rho=2$ is presented in Fig. \ref{f:structures}, lower right.

The special case $\rho=2$ ($J_2/J_1=2$) for $h > 0$ has been solved in Ref. \onlinecite{moliner09prb79:144401}. 
In this case the Hamiltonian (\ref{hamilt}) possesses an additional degeneracy, and there is an infinite number
of spin structures which share both total energy and the magnetization with the umbrella structure introduced below.
The generic case $h>0$ is a bit more complicated. First we introduce several candidate spin structures which
correspond to the local extrema of the total energy and
present the phase diagram of the SSL in the $(\rho,h)$ plane. In the next section we prove rigorously
that the three structures considered (spin-flop, spin-flip and umbrella) are indeed the lowest-energy 
structures in the respective regions of the $(\rho,h)$ plane.

UUD structure, shown in  Fig. \ref{f:structures}, has
energy and magnetization per site:
\be
\label{e-uud}
\epsilon_{UUD}= -\frac 23 - \frac {\rho}6 - \frac h3, \qquad M_{UUD}=1/3.
\ee
Here and in the following we define $M$ as the magnetization per lattice site,
so that $M_s=1$. The first term in Eq. (\ref{e-uud}) is calculated by including
$16$ edges ($J_1$ bonds) within the unit cell of the UUD structure with factor $1$,
and $16$ edges which cross the unit cell boundary with factor $1/2$. The total contribution
of these terms to the unit cell energy is equal to $-8$, which gives per site contribution of
$-2/3$ when divided by the number of sites. 

In the dimer structure every $J_2$ bond connects two oppositely alligned spins (e.g. $+\vz$ and $-\vz$).
This can be done in an infinite number of ways as each dimer can be oriented independently from
all others. This structure satisfies all $J_2$ bonds and its energy is
\be
\label{e-dimer}
\epsilon_{Dimer}= - \frac {\rho}2, \qquad M_{Dimer}=0.
\ee
As we will see below, the dimer structure is never a ground state
of the isotropic classical Heisenberg SSL, except in the limit $J_1=0$ ($\rho=\infty$).
It is important for the anisotropic Heisenberg and Ising SSL, however, and it is
also realized for finite values of $\rho$ in a quantum Heisenberg SSL \cite{shastry81physicabc108:1069}.
 
The spin-flop structure is the Neel structure with all spins tilted by the angle
$\theta^{\prime}=\pi/2-\theta=\sin^{-1} \lk {h}/{8} \rk$ out of the $xy$ plane.
This solution exists for $h < 8$, and its energy
and magnetization are given by
\be
\label{e-flop}
\epsilon_{Flop}=-2+\frac{\rho}2 - \frac{h^2}{16} , \qquad M_{Flop}=\cos(\theta)=\frac{h}{8}.
\ee
The spin-flip (ferromagnetic) structure (Fig. \ref{f:structures}, upper right)
 has all spins aligned along the magnetic field. It's energy is
\be
\label{e-flip}
\epsilon_{Flip}=2 + \frac{\rho}2 - h, \qquad M_{Flip}=1.
\ee
At $h=8$ the spin-flop structure turns into the spin-flip structure in a continuous fashion.
Since $M(h)$ is continuous at the point $h=8$, but $\partial M(h)/\partial h$ is not,
this is a second-order phase transition.

\begin{figure}
\includegraphics[scale=0.32]{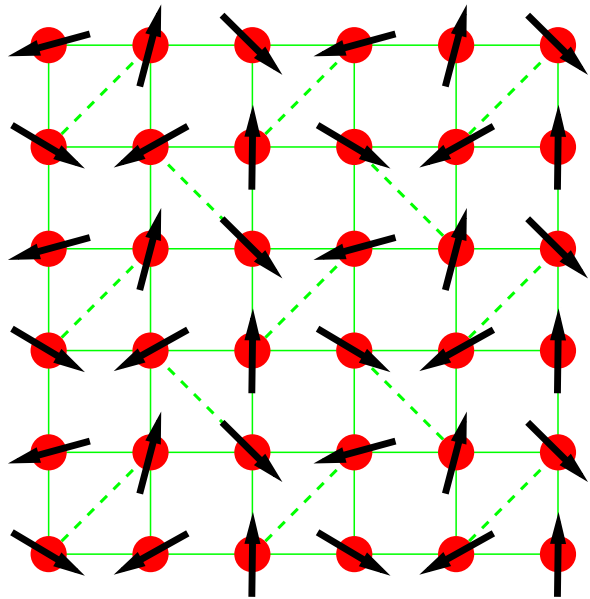}
\caption{\label{f:umb1} (Color online) Umbrella structure for $\rho=1.5$, $h=3$ calculated numerically on the 6 x 6 lattice
with periodic boundary conditions. The arrows show the $xy$ components of spins. The red (gray) circles
show the $z$ components of spins (all equal for the umbrella structure).}
\end{figure}

\begin{figure}
\includegraphics[scale=0.16]{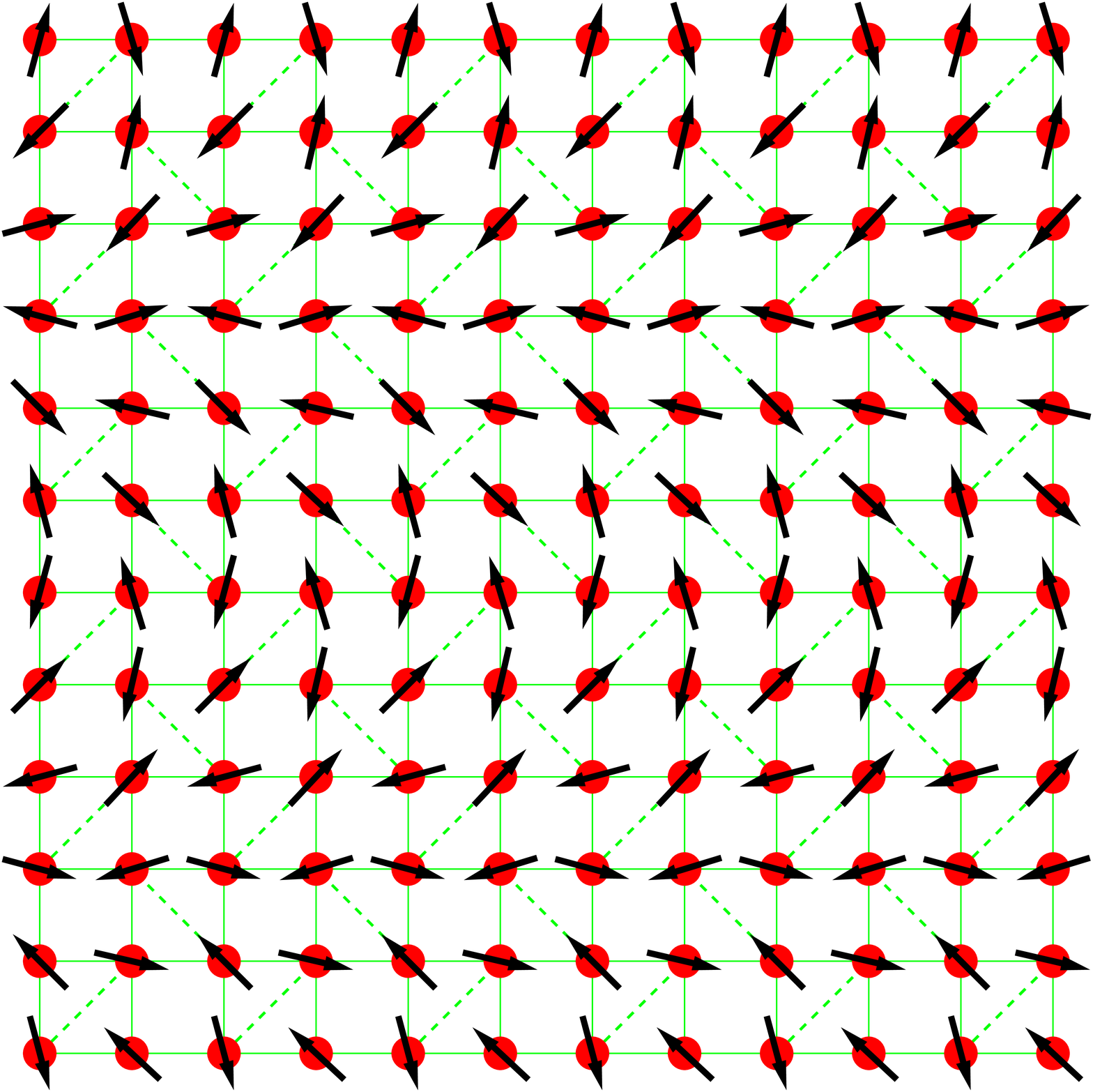}
\caption{\label{f:umb2} (Color online) Umbrella structure for $\rho=1.2$, $h=3$ calculated numerically on the 12 x 12 latice
with periodic boundary conditions. The arrows show the $xy$ components of spins. The red (gray) circles
show the $z$ components of spins (all equal in this case).}
\end{figure}

Another possible structure is the umbrella structure, proposed in Ref. \onlinecite{moliner09prb79:144401}.
It is essentially a tilted spiral structure. In the umbrella structure the spherical angles $\theta_i$ of all
spins are equal, and the angles $\phi_i$ are distributed like in the spiral structure above. The
energy for given $\theta, \triangle\phi$ is
\begin{multline}
\epsilon(\theta, \triangle\phi) = 2 \lk -\sin^2 \theta \cos(\triangle\phi) + \cos^2 \theta \rk + \\
\frac {\rho}2 \lk \sin^2 \theta \cos (2\triangle\phi) + \cos^2 \theta \rk - h \cos \theta.
\end{multline}
Minimization with respect to $\theta, \triangle\phi$ gives
 $\triangle\phi=\cos^{-1} (1/\rho)$ (it does not depend on $h$), and
\be
\label{m-umb}
M_{Umb}=\cos \theta = \frac{h \rho}{2\lk \rho+1 \rk^2},
\ee
\be
\label{e-umb}
\epsilon_{Umb}= -\frac{1}{\rho} - \frac {\rho}2 - \frac{h^2 \rho}{4\lk \rho+1 \rk^2}.
\ee
The umbrella structure exists for $\rho>1$ and $h<h_{max}=2\lk \rho + 1 \rk^2/\rho$, has energy lower than the spin-flop one,
and turns into the spin-flop structure at $\rho=1$. At $h=h_{max}$ it becomes the
spin-flip structure. Both phase transitions are of the second order. In Fig. \ref{f:umb1} and Fig. \ref{f:umb2}
numerically calculated umbrella structures for $(\rho,h)=(1.5,3)$ and $(1.2,3)$ respectively are presented
(See section \ref{sec:num} below for details).

\begin{figure}
\includegraphics[scale=0.33]{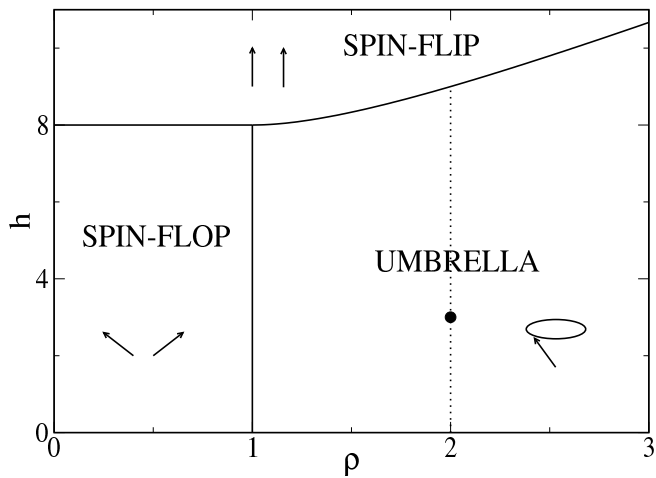}
\caption{\label{f:iso} $T=0$ phase diagram of the classical SSL. The solid lines correspond to second order
phase transitions. The dotted line marks the special degenerate case $J_2/J_1=2$.
The large dot at $(\rho,h)=(2,3)$ marks the single point in the $(\rho,h)$ plane where
the UUD structure can exist.}
\end{figure}

The phase diagram of the 
classical isotropic SSL is shown in Fig. \ref{f:iso}. The solid lines mark the lines of the 
second order phase transitions. For $\rho<1$
SSL behaves exactly like a regular Neel antiferromagnet, while for $\rho > 1$ the spin-flop phase is replaced 
by the umbrella phase. The dotted line denotes the special degenerate case $\rho=2$. 
Magnetization curve $M(h)$ are linear until saturation, and there is no magnetization
plateaus for the isotropic Heisenberg SSL. Note that the $1/3$ pseudo-plateau observed in Ref. \onlinecite{moliner09prb79:144401}
was a finite-temperature effect. The phase diagram in Fig. \ref{f:iso} has been constructed
by comparing energies of different spin structures introduced in this section and selecting
the one with the lowest energy for a given $(\rho,h)$. In order to prove that this phase diagram is
indeed correct, we have to show that there are no other spin structures with lower energy. This is 
done in the next section.

\section{\label{sec:sol}Exact ground state}
In order to prove that the three spin structures introduced in the previous section
(spin-flop, spin-flip and umbrella) are 
indeed the ground state structures for respective $(\rho,h)$ we use the method of decomposing the Hamiltonian
into overlapping elementary clusters, which has been used previously in, e.g.,
Refs. \onlinecite{dublenych12prl109:167202,shastry81physicabc108:1069,moliner09prb79:144401}
\be
\label{ham_sum}
\widetilde{\mathcal{H}} = \sum_{\triangle} \mathcal{H}_{\triangle}.
\ee
For SSL such elementary cluster has the shape of the right triangle,
highlighted in red in Fig. \ref{f:ssl}.
Each triangle includes one spin $\vS_0$ at the right angle, two spins ($\vS_1$ and $\vS_2$) at the
$45^{\circ}$ angles; two $J_1$ bonds, and one $J_2$ bond shared by two triangles. The number of triangles
on the lattice is the same as the number of lattice sites, as each triangle includes 3 spins and each
spin $\vS_i$ is a part of 3 triangles, acting the roles of $\vS_0$, $\vS_1$ and $\vS_2$ in turn.
The Hamiltonian of a triangle is
\be
\label{ham_trig}
\mathcal{H}_{\triangle} = \vS_0  \cdot \lk \vS_1 + \vS_2\rk + \frac {\rho}2 \vS_1  \cdot \vS_2 -
\vh  \cdot \lkv \alpha \vS_0 + \frac {1-\alpha}2 \lk \vS_1 + \vS_2 \rk \rkv,
\ee
where $\alpha$ is an arbitrary real number. It corresponds to an arbitrary way in which the
term $-\vh  \cdot \vS_i$ of the original Hamiltonian can be divided between three different triangles
which include the site $\vS_i$. While the triangle Hamiltonian $\mathcal{H}_{\triangle}$ depends on $\alpha$,
the lattice Hamiltonian $\widetilde{\mathcal{H}}$ does not, as all $\alpha$-dependent terms cancel each other 
upon summation in Eq. (\ref{ham_sum}).

 The triangle Hamiltonian is invariant under three basic symmetry
operations: simultaneous rotation of all spins around $z$-axis, reflection of all spins in the
$xz$ plane, and interchange $\vS_1 \leftrightarrow \vS_2$, where
$z$-axis is parallel to $\vh$, and $x$-axis is an arbitrary axis perpendicular to $z$-axis.
 These operations generate a symmetry group, which also
includes such operations as reflection in the $yz$ plane (or any other plane containing $z$-axis) and
inversion of the $xy$ components of all spins $\vS_i^x \to -\vS_i^x$, $\vS_i^y \to -\vS_i^y$ ($i=0,1,2$).
This symmetry leads to the
degeneracy of most energy levels $\epsilon(\alpha,\vS_0$, $\vS_1$, $\vS_2)$, as
the only configurations invariant under the symmetry group of $\mathcal{H}_{\triangle}$ are
collinear ones with $\vS_0=\pm \vz$ and $\vS_1=\vS_2=\pm \vz$.
At the special point $\rho=2$ there is an additional symmetry operation
$\vS_0 \leftrightarrow \vS_1$ (or $\vS_0 \leftrightarrow \vS_2$).

The triangle Hamiltonian has an $\alpha$-dependent
ground state energy $\epsilon_0(\alpha)$,
thus for every possible configuration of three unit vectors
$\vS_0$, $\vS_1$, and $\vS_2$ the inequality
\be
\epsilon_{\triangle}(\vS_0,\vS_1, \vS_2, \alpha) \ge \epsilon_0(\alpha)
\ee
holds true. The equality is achieved for a possibly degenerate
ground state configuration of the triangle Hamiltonian.
There is also an inequality for the lattice energy,
which holds true for all values of $\alpha$
\be
\label{ineq}
N \epsilon \lfi S_i \rfi = \sum_{\triangle} \epsilon_{\triangle} \ge \sum_{\triangle} \epsilon_0(\alpha) =
N \epsilon_0(\alpha),
\ee
or $\epsilon \lfi S_i \rfi \ge \epsilon_0(\alpha)$, 
where $N$ is the number of lattice sites, and $\epsilon \lfi S_i \rfi$ is the energy per site.
The equality here is possible only for a spin structure $\lfi \vS_i \rfi$ that minimizes 
the energy of each triangle simultaneously. Here and in the following we use the word "structure"
for spin structures $\lfi \vS_i \rfi$ on the lattice, and the word "configuration" for
configurations $\vS_0$, $\vS_1$, and $\vS_2$ of the three spins of a triangle.
According to (\ref{ineq}), for every other spin structure
$\lfi \vS_i^{\prime} \rfi$ one can write an inequality
$\epsilon \lfi S_i^{\prime} \rfi \ge \epsilon_0(\alpha) = \epsilon \lfi S_i \rfi$,
which proves that the structure $\lfi S_i \rfi$ is indeed the ground state of
the lattice Hamiltonian (\ref{hamilt2}), or, in general, one of the degenerate
ground state structures. In other words, in order to find the ground state of the
lattice Hamiltonian (\ref{hamilt2}) for a given $(\rho,h)$,
we have to construct a 
lattice spin structure $\lfi S_i \rfi$ from the ground state configuration 
$(\vS_0,\vS_1,\vS_2)$ of a triangle (let us call it "brick"), or from a set of such bricks
in case of degeneracy. In case of a degenerate ground state, it is important to note
that any possible lattice structure constructed from bricks is a ground state structure,
and, vice versa, any possible ground state structure can be constructed from bricks
(which can be seen from the fact that for any ground state structure the energy
of each triangle is equal to $\epsilon_0$).
 
The question is whether it is possible to construct a lattice structure from a given set of bricks,
which includes all possible realization of the degenerate ground state of $\mathcal{H}_{\triangle}$,
 spawned by the symmetry group of $\mathcal{H}_{\triangle}$.
The problem is not trivial, since each spin is a part of three
different triangles, so the bricks must match each other perfectly.
This is obviously not possible for arbitrary $\alpha$, however, 
as we are going to see below, such construction can indeed be performed for the right
choice of $\alpha$. The required values are $\alpha=1/2$ for $\rho \le 1$, and
$\alpha = 1/(\rho+1)$ for $\rho>1$, respectively. For these values of $\alpha$
we can construct the spin-flop, spin-flip and umbrella structures from the
ground state configurations of $\epsilon_{\triangle}$ and confirm the
phase diagram of Fig. \ref{f:iso}. Contrary to the case of the Ising SSL \cite{dublenych12prl109:167202},
the phase diagram of the classical Heisenberg SSL does not have the convexity property,
so the search for ground state must be performed for arbitrary $(\rho,h)$,
rather than for a finite number of special points.

\begin{table}
\begin{tabular}{|c|c|c|}
\hline  & Energy & Domain \\
\hline Flip & $2+ \rho/2 \mp h$ & \\
Neel & $-2+\rho/2 \mp h (1-2\alpha)$ & \\
Dimer & $-\rho/2 \mp h \alpha$ &  \\ 
Umb & $-1/\rho - \rho/2 - \frac{h^2 \rho}{4\lk \rho+1 \rk^2}$ & 
$\rho \ge 1, \: h < \frac{2(\rho+1)^2}{\rho} $ \\ 
$Y_1$ & $-\frac 1{\rho} - \frac {\rho}2 + 
h \lk \alpha - \frac{\alpha^{\prime}}{\rho}\rk - \frac{{\alpha^{\prime}}^2h^2}{4\rho}$ &
$\frac{\alpha^{\prime}h}{2} < \rho-1 $ \\
$Y_2$ & $-\frac 1{\rho} - \frac {\rho}2 - 
h \lk \alpha - \frac{\alpha^{\prime}}{\rho}\rk - \frac{{\alpha^{\prime}}^2h^2}{4\rho}$ &
$1-\rho < \frac{\alpha^{\prime}h}{2} < 1 + \rho $ \\
Flop & $\frac{\rho}2 - \frac{\alpha(1-\alpha)h^2}4 -
 \lk \frac{\alpha^{\prime}}{\alpha} + \frac{\alpha}{\alpha^{\prime}} \rk$ & 
$h< \frac{2}{\alpha\alpha^{\prime}}$\\
\hline
\end{tabular}
\caption{\label{tab:eps} Energies $\epsilon(\alpha)$ and domains of existence
of different configurations of a triangle,
which are energy extrema of $\mathcal{H}_{\triangle}$.
We use the definition $\alpha^{\prime} \equiv 1 - \alpha$.
 The umbrella-like configuration
is only defined for $\alpha=1/(\rho+1)$.}
\end{table}

For simplicity we assume that $h>0$, so that the field $\vh$ provides a fixed direction $\vz$.
The rather trivial case $h=0$ (solved in Ref. \onlinecite{shastry81physicabc108:1069}
by the same method) has been discussed in the previous section.
Let us find all possible steady states (energy extrema) of the triangle Hamiltonian (\ref{ham_trig})
and their energies (listed in Table \ref{tab:eps}).  They can be
found from the system of three vector equations
\be
\frac {\partial}{\partial \vS_j} \lk \mathcal{H}_{\triangle} - 
\frac 12 \sum_i \xi_i \vS_i  \cdot \vS_i \rk =0, \quad i,j=0,1,2
\ee
for the three unit vectors $\vS_0, \vS_1, \vS_2$:
\be
\label{veq1}
\frac {\partial \mathcal{H}_{\triangle}} {\partial \vS_1} = \vS_0 + \frac {\rho}2 \vS_2 -
\frac {1-\alpha}2 \vh = \xi_1 \vS_1,
\ee
\be
\label{veq2}
\frac {\partial \mathcal{H}_{\triangle}} {\partial \vS_2} = \vS_0 + \frac {\rho}2 \vS_1 -
\frac {1-\alpha}2 \vh = \xi_2 \vS_2, \\
\ee
\be
\label{veq0}
\frac {\partial \mathcal{H}_{\triangle}} {\partial \vS_0} = \vS_1 + \vS_2  
- \alpha \vh = \xi_0 \vS_0,
\ee
where $\xi_i$ are the three real Lagrange multipliers
used to enforce the conditions $\vS_i  \cdot  \vS_i = 1$.
They are nonpositive for local energy minima,
and nonnegative for local maxima.

First, let us consider collinear ($\vS_i = \pm \vz$) 
solutions of Eqs. (\ref{veq1})--(\ref{veq0}). Any collinear configuration is a solution.
The energies of two spin-flip-like configuration ($\vS_i=\pm\vz$) are
$\epsilon_{Flip1,2}=2+ \rho/2 \mp h$ and do not depend on $\alpha$. The two Neel-like
configurations ($\vS_1=\vS_2=-\vS_0=\pm\vz$) have energies
$\epsilon_{Neel1,2}=-2+\rho/2 \mp h (1-2\alpha)$. Finally, the two dimer-like
configurations ($\vS_1=-\vS_2=\vz, \vS_0=\pm \vz$) have energies 
$\epsilon_{Dimer1,2}=-\rho/2 \mp h \alpha$. 
Now let us find all noncollinear solutions
of Eqs. (\ref{veq1})--(\ref{veq0}). Subtracting first two equations gives
\be
\vS_1 \lk \xi_1 + \frac {\rho}2 \rk = \vS_2 \lk \xi_2 + \frac {\rho}2 \rk.
\ee
It means that either $\vS_1$ and $\vS_2$ are collinear, or that both expressions
in parentheses are equal to zero. There are 3 possible cases
\[
\text{\textbf{Case 1:}} \quad \xi_1 = \xi_2 = - \rho/2,
\]
\[
\text{\textbf{Case 2:}} \quad \vS_1 = - \vS_2,
\]
\[
\text{\textbf{Case 3:}} \quad \vS_1 = \vS_2.
\]
Below we shall consider all 3 cases in detail.

\textbf{Case 1:} $\xi_1 = \xi_2 = - \rho/2$. We introduce a new variable $\vS \equiv (\vS_1+\vS_2)/2$,
and note that $0 \le \lab \vS \rab \le 1$, but $\lab \vS_0 \rab=1$.
The equations (\ref{veq1})--(\ref{veq0}) become
\be
\vS_0 + \rho \vS  = \vh (1-\alpha)/2,
\ee
\be
2 \vS - \alpha \vh = \xi_0 \vS_0.
\ee
Excluding $\vS_0$ gives
\be
\vS \lk 2 + \xi_0 \rho \rk = \vh \lkv \alpha + \xi_0 (1-\alpha)/2\rkv.
\ee
There are again two possible cases. 

\textbf{Case 1.1:} $2 + \xi_0 \rho = \alpha + \xi_0 (1-\alpha)/2 = 0$. 
This is only possible
for $\alpha = 1/(\rho+1)$ (for $\alpha=1/2$, $\rho<1$ there are
no solutions of this kind), which shows that it is indeed the correct
value of $\alpha$ for $\rho>1$, since no other value of $\alpha$ can
produce the umbrella phase. For this value of $\alpha$ there is
an entire family of the degenerate "umbrella-like" solutions of the form
\be
\label{umb_s0s}
\vS_0 + \rho \vS = \frac {\rho \vh}{2(\rho+1)}
\ee
for $h<h_{max}=2\lk \rho + 1 \rk^2/\rho$
with the energy
\be
\epsilon_{Umb}=-\frac{1}{\rho} - \frac {\rho}2 - \frac{h^2 \rho}{4\lk \rho+1 \rk^2}.
\ee
This energy is found from Eqs. (\ref{ham_trig}), (\ref{umb_s0s}) using the identities
\be
\vS_1 \cdot \vS_2 = 2 \vS^2 -1,
\ee
\be
2 \vS_0 \cdot \vS + \rho \vS^2 = \frac 1\rho \lkv (\vS_0+\rho \vS)^2 -1 \rkv,
\ee
and also the fact that
\be
\lkv \alpha \vS_0 + \frac {1-\alpha}2 \lk \vS_1 + \vS_2 \rk \rkv=
 \frac 1{\rho+1} \lk \vS_0 + \rho \vS\rk
\ee
for $\alpha=1/(\rho+1)$.

Any structure constructed from umbrella-like bricks has magnetization given by
Eq. (\ref{m-umb}). This can be seen by averaging Eq. (\ref{umb_s0s}) over all triangles
and using $\left< \vS_0 \right> = \left< \vS \right> = M$.
For case $\rho=2$, the degeneracy of the umbrella-like brick 
is fully preserved at the lattice level \cite{moliner09prb79:144401},
namely, every configuration with $\vS_0 + \vS_1 + \vS_2 = {\vh}/3$
can form a lattice structure consisting of three types of sites (for example, 
ordered as in Fig. \ref{f:structures}, lower right), since $\vS_0$, $\vS_1$, and
$\vS_2$ can be interchanged freely. In particular, the UUD structure
can be realized at a single point $(\rho,h)=(2,3)$ (large dot in Fig. \ref{f:iso}).
For $\rho \ne 2$ the requirement of bricks matching each other partly
lifts the degeneracy. The umbrella structure introduced above, for which
$\vS_i  \cdot \vh = \cos \theta$ is equal for all spins, is
a possible way to match bricks on the lattice. In our numerical simulations
(see the next section) we found no other structures degenerate with the umbrella
one for $\rho \ne 2$, but we were not able to prove their absence rigorously.

\textbf{Case 1.2:} $\vS$, $\vS_0$ and $\vh$ are collinear (but $\vS_{1,2}$ and $\vh$ are not). 
This gives up to two Y-like configurations, with $\vS \parallel \vz$ and $\vS_0 = \pm \vz$.
The $Y_1$ configuration with $\vS_0=-\vz$ has the energy
\be
\epsilon_{Y_1}(\alpha)= -\frac 1{\rho} - \frac {\rho}2 + 
h \lk \alpha - \frac{1-\alpha}{\rho}\rk - \frac{(1-\alpha)^2h^2}{4\rho}
\ee
and it exists for $h<2 (\rho-1)/(1-\alpha)$.
The $Y_2$ configuration with $\vS_0=+\vz$ has the energy
\be
\epsilon_{Y_2}(\alpha)= -\frac 1{\rho} - \frac {\rho}2 - 
h \lk \alpha - \frac{1-\alpha}{\rho}\rk - \frac{(1-\alpha)^2h^2}{4\rho}
\ee
and it exists for
$1-\rho < (1-\alpha)h/2 < 1 + \rho$.

For $\alpha = 1/(\rho+1), \rho>1$ the $Y_{1,2}$ solutions 
are just two special cases of the umbrella-like solution introduced above.
For $\alpha=1/2, \rho \le 1$ there is a single $Y_2$ solution with the energy
\be
\epsilon_{Y_2}=-\frac{1}{\rho} - \frac {\rho}2 - \frac{h (\rho-1)}{2\rho} - \frac{h^2}{4\rho}.
\ee

\textbf{Case 2:} $\vS_1=-\vS_2$. The dimer-like collinear solutions belong to this case, but there are no
noncollinear solutions.

\textbf{Case 3:} $\vS_1=\vS_2=\vS$ and $\xi_1=\xi_2=\xi$.
The equations (\ref{veq1})--(\ref{veq0}) take the form
\be
\vS_0 + \lk \frac{\rho}2 - \xi \rk \vS = \frac{1-\alpha}{2} \vh
\ee
\be
2\vS - \alpha \vh = \xi_0 \vS_0,
\ee
or, after excluding $\vS_0$,
\be
\vh \lkv \alpha + \frac{\xi_0 (1-\alpha)}{2}\rkv =
\vS \lkv 2 - \xi_0 \lk \xi - \frac{\rho}{2} \rk \rkv.
\ee
As before, it can mean either that $\vS$ is collinear with $\vh$ (which leads to
collinear spin-flip and Neel-like solutions), or that both expressions 
in the square brackets are equal to zero, which leads to
\be
\xi_0 = - \frac{2\alpha}{1-\alpha}, \quad 
\xi = \frac{\rho}{2} - \frac{1-\alpha}{\alpha},
\ee
and
\be
\vS + \vS_0 \frac{\alpha}{1-\alpha} = \frac{\alpha}{2} \vh, 
\ee
which is the spin-flop-like solution with the energy
\be
\epsilon_{Flop}(\alpha)=\frac {\rho}2 - \frac{\alpha(1-\alpha)}{4}h^2 - \frac{1-\alpha}{\alpha} - 
\frac{\alpha}{1-\alpha}.
\ee
It exists for $h< 2 / (\alpha(1-\alpha))$.
For $\alpha = 1/(\rho+1), \rho>1$ this is again a special case of the
umbrella-like solution. For $\alpha=1/2, \rho \le 1$ this is the
spin-flop solution ($\vS + \vS_0 = \vh/4$) with the energy
$\epsilon_{Flop}=-2 + \rho/2 - {h^2}/{16}$. The spin-flop-like
bricks only match if $\vS  \cdot \vh = \vS_0  \cdot \vh$, i.e. that all spins
are tilted by the same angle $\theta^{\prime}$ relative to the $xy$-plane.
This condition gives $\alpha=1/2$ as the only value of $\alpha$ for which
the construction of the spin-flop lattice structure is possible.

Some of the configurations corresponding to the energy extrema of 
$\mathcal{H}_{\triangle}$ do not match and thus cannot form a spin
structure on the lattice. The ones that do are: spin-flop and Neel (for $\alpha=1/2$),
umbrella-like (for $\alpha=1/(\rho+1)$), dimer (for $\alpha=0$), and
spin-flip (for any $\alpha$). The direct comparison of their energies
(listed in Table \ref{tab:eps}) for respective values 
of $\alpha$ shows that spin-flop, umbrella-like and
spin-flip solutions indeed minimize the triangle Hamiltonian (\ref{ham_trig})
in the three respective regions of Fig. \ref{f:iso}.
The Neel structure is only realized for $h=0, \: \rho \le 1$.
The dimer structure does not exist for the isotropic classical Heisenberg
SSL. This confirms the phase diagram shown in Fig. \ref{f:iso}.

\section{\label{sec:num} Numerical simulations and the effect of anisotropy}

\begin{figure}
\includegraphics[scale=0.31]{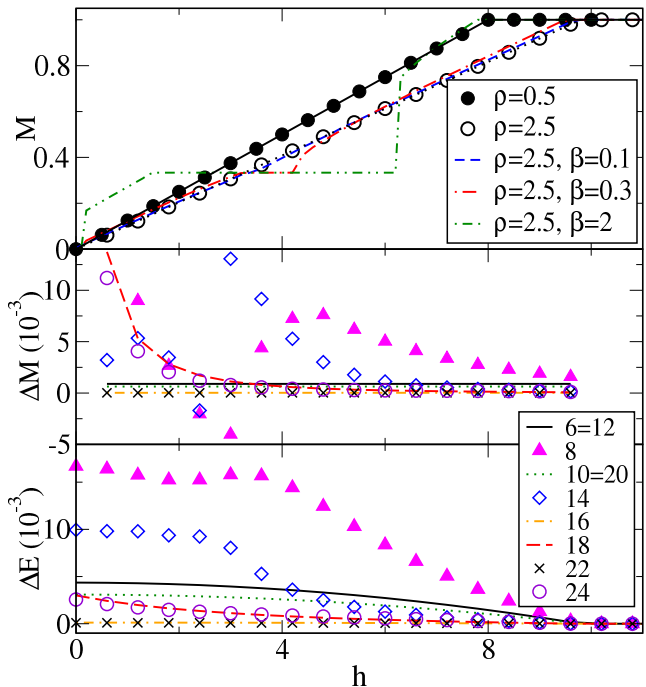}
\caption{\label{f:mag-iso} (Color online) Upper panel: Magnetization curves $M(h)$ for the classical SSL. 
$\rho=0.5$, no anisotropy: exact result (solid black line), numerical data (solid circles).
$\rho=2.5$, no anisotropy: exact result (dotted black line), numerical data (empty circles).
$\rho=2.5$, with anisotropy: $\beta=0.1$ (blue dashed curve), $\beta=0.3$ (red dash-dot curve),
$\beta=2$ (green dash-dot-dot curve).
All numerical calculations have been performed on the 12 x 12 lattice with periodic
boundary conditions (except for $\beta=2$, for which 6 x 2 lattice has been used).
The calculations with anisotropy used a uniform $h$-grid of 61 points
ranged between $h=0$ and $h=12$ (121 $h$-points for $\beta=2$). 
Middle and lower panels: differences $\triangle M(h) \equiv (M_{calc}(h)-M_{exact}(h))/M_{exact}(h)$
and $\triangle E(h) \equiv (E_{calc}(h)-E_{exact}(h))/J_1$ between calculated (on an $n$ x $n$ lattice) and exact 
magnetizations and energies respectively for $\rho=2.5$, $\beta=0$, and different values of $n$.}
\end{figure}

In order to give an independent check of our exact results we performed a series 
of numerical simulations, calculating the ground state of lattice Hamiltonian 
$\widetilde{\mathcal{H}}$ for different values of $\rho$, $h$. We have also examined
the effect of uniaxial anisotropy by adding the term 
\be
\mathcal{H}_A = \frac B2 \sum_i \lk 1 - (S_i^z)^2 \rk 
\ee
to the Hamiltonian $\mathcal{H}$, or, equivalently, adding the term
\be
\widetilde{\mathcal{H}}_A = \frac{\beta}2 \sum_i \lk 1 - (S_i^z)^2 \rk
\ee
to $\widetilde{\mathcal{H}}$,
where $\beta \equiv B/J_1$ is the anisotropy constant.

The minimum of $\widetilde{\mathcal{H}}$ was found by a discrete
micromagnetic simulation with only the Gilbert damping term included. We considered the 
system of equations
\be
\frac{d\vS_i}{dt} = - \lambda \lkv \frac{\partial \widetilde{\mathcal{H}}}{\partial \vS_i} - 
\vS_i \lk \vS_i \cdot \frac{\partial \widetilde{\mathcal{H}}}{\partial \vS_i} \rk \rkv,
\ee
where $\lambda>0$ is the damping parameter,
for each spin $\vS_i$ of the lattice; and solved it  
using the first-order Runge--Kutta method in spherical coordinates
with our own computer code. This method
decreases the energy of the system on each step (provided that $\lambda$ is small enough)
eventually finding a (local) minimum. It can be viewed as a variation of the steepest descend method.
Unlike previously used Monte Carlo methods, our method looks for the ground state of the system avoiding
any finite-temperature effects. We have used square cells of different size
(usually 6 x 6, 12 x 12, or 24 x 24)
with periodic boundary conditions. In order to minimize the probability of finding a local energy minimum
instead of the ground state,
each simulations was performed 50 times with different random initial conditions, and the 
result with the lowest energy was chosen.

\begin{figure}
\includegraphics[scale=0.32]{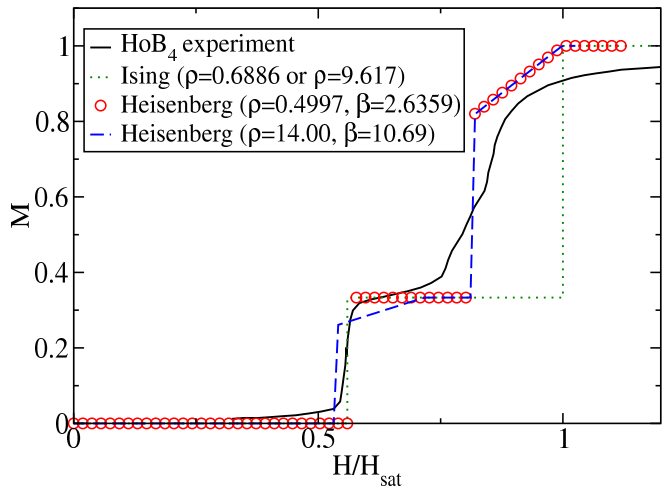}
\caption{\label{f:hob4} (Color online) Experimental and theoretical magnetization curves
$M(H/H_{sat})$ for HoB$_4$: experimental data from Ref. \onlinecite{matas10jpcs200:032041} (solid black curve);
Ising data for $\rho=0.6886$ or $9.617$ (dotted green line); Heisenberg data for $\rho=0.4997$, $\beta=2.6359$,
61 $h$-points (red circles) and $\rho=14.0$, $\beta=10.69$,
111 $h$-points (dashed blue line). $H_{sat}$ is the saturation field.
Heisenberg data has been calculated on the 6 x 2 lattice.}
\end{figure}

A number of simulations for different values of $\rho$, $h$ have been performed.
Our results fully confirm the phase diagram of Fig. \ref{f:iso}, in particular,
the second order phase transitions and the additional degeneracy for $\rho=2$ are
clearly seen in our calculations. We have also applied the same method to the 
triangle Hamiltonian $\mathcal{H}_{\triangle}$ and checked numerically the validity
of the results of the previous section. Two typical 
magnetization curves $M(h)$ are presented in Fig. \ref{f:mag-iso} (upper panel) for 
$\rho=0.5$ and $\rho=2.5$. The numerical results
are plotted as circles in this figure, while the analytical formulae (\ref{e-neel})--(\ref{m-umb}) are
presented as lines. For $\rho=0.5$ the system goes through the sequence of the Neel--spin-flop--spin-flip
structures, all of them being periodic with 4 atoms per unit cell, exactly as predicted by our analytical
treatment above. The $M(h)$ curves in fact do not depend on $\rho$ at all as long as $\rho \le 1$.
For $\rho > 1$ the ground state is the umbrella structure, which is 
in general incommensurate with the lattice.
The numerical calculations with a finite lattice size cannot reproduce this structure exactly,
of course. Two examples of the calculated umbrella structures with periodicity forced by the lattice size
are presented in Fig. \ref{f:umb1} and Fig. \ref{f:umb2} for $(\rho,h)=(1.5,3)$ and $(1.2,3)$ respectively.

In order to analyze the dependence of the calculated magnetization and
energy of the umbrella structure on the lattice size, we have calculated the $M(h)$ and $E(h)$ curves for 
$\rho=2.5$ using $n$ x $n$ square lattices for different values of $n$ and compared 
them to the exact results. The relative magnetization difference 
$\triangle M(h) \equiv (M_{calc}(h)-M_{exact}(h))/M_{exact}(h)$ and the absolute
energy difference  $\triangle E(h) \equiv (E_{calc}(h)-E_{exact}(h))/J_1$ are presented
in Fig. \ref{f:mag-iso}, middle and lower panels respectively, for $n$ ranging from 6 to 24.
The results for $n=6$ and $n=12$ are identical, as are the ones for $n=10$ and $n=20$.
The general trend of the convergence of the calculated $M$ and $E$ to the exact values
upon the increase of $n$ is clearly seen, although the process of the convergence
is far from steady.
The best results are obtained for the 16 x 16 and 22 x 22 lattices, while
the results for the 8 x 8 and 14 x 14 lattices are the worst. This stems from the fact that 
the elementary angle $\triangle\phi=\cos^{-1} (1/\rho)$ of the umbrella phase
is approximated on a $n$ x $n$ lattice by $\frac mn \cdot 360^{\circ}$, with integer $m$.
For $\rho=2.5$ this angle is $\triangle\phi \approx 66.422^{\circ}$ and it
is close to $\frac 3{16} \cdot 360^{\circ}$ and $\frac 4{22} \cdot 360^{\circ}$.
The worst results are generally obtained when $\triangle\phi$ is close to $(m+1/2)/n \cdot 360^{\circ}$,
as is the case for $n=8$ and $n=14$. With the increase of $n$ the rational approximations
$\frac mn \cdot 360^{\circ}$ converge to the exact $\triangle\phi$.
As long as the lattice size is $6 \times 6$ or larger, the differences $\triangle M$ and
$\triangle E$  are smaller or of the order of $10^{-2}$, a difference too small to be
seen in the scale of Fig. \ref{f:mag-iso} (upper panel). For $n=6$, $\triangle M$ is
of the order of $9 \cdot 10^{-4}$. For $n=24$ the calculated $M$ is actually worse 
compared to $n=6$ for very small $h$, however, for larger values of $h$, the calculated
magnetization is better for $n=24$; and, more importantly, the calculated energy for $n=24$ is
lower and therefore $\triangle E$ is smaller compared to $n=6$ for all values of $h$.
We have also calculated the $M(h)$ and $E(h)$ curves for the $n$ x $2$ lattices,
where $n=6, 8, \ldots 24$. The results (not shown) are identical to the ones
obtained for the $n$ x $n$ lattices. The reason for this is that the typical 
umbrella structure (see e.g. Fig. \ref{f:umb1} and Fig. \ref{f:umb2}) has period 2
in one of the directions ($x$ or $y$).
 
If the magnetic anisotropy $\beta$ is switched on, the single UUD point $(\rho,h)=(2,3)$
(large dot in Fig. \ref{f:iso}) expands into
a finite region of the UUD phase. The $M(h)$ curves for $\rho \ne 2$ first get a noticeable kink at around
$M=1/3$ (Fig. \ref{f:mag-iso}, upper panel, dashed blue curve), which eventually turns into
the $M=1/3$ UUD plateau when the anisotropy constant is increased
(Fig. \ref{f:mag-iso}, upper panel, dashed-dot red curve). For $\rho = 2$ the $M=1/3$ step appears for
any finite value of $\beta$. Our $M(h)$ curves are very similar to the 
$M(h)$ curves of Ref. \onlinecite{qin11jap109:07E103}, which were obtained using Monte Carlo
method for $\rho=2$. However, we observed the onset of the $1/3$ plateau at a finite
value of $\beta$, which was not seen in Ref. \onlinecite{qin11jap109:07E103} due
to the choice of a non-arbitrary point $\rho=2$. No fractional plateaus other than $1/3$ appear
in our calculations. Apart from the UUD structure, the anisotropy also stabilizes other collinear structures.
The stabilization of the spin-flip structure leads to the decrease of the saturation field with increasing
$\beta$. A region of the Neel phase appears in the $(\rho,h)$ plane, leading to
the $M=0$ plateau in the $M(h)$ curve and the spin-flop transition at a final $h$. For $\rho \le 1$ this step appears 
for any finite $\beta$, while for $1<\rho<2$ it appears for $\beta$ above a certain threshold.
For $\rho>2$ and large enough $\beta$ the dimer structure is the ground state at small $h$, which also leads to
a $M=0$ plateau (Fig. \ref{f:mag-iso}, upper panel, dash-dot-dot green curve). 
The $M=0$ plateau cannot exist for $\rho=2$, not even in the Ising model  \cite{dublenych12prl109:167202}.
Futher examples of the $M(h)$ curves with the $M=0$ plateau can be seen in Fig. \ref{f:hob4}, see
the discussion below.

Let us compare our results to the experimental $M(H)$ curve of HoB$_4$
(Ref. \onlinecite{matas10jpcs200:032041}),
which contains a single $1/3$ fractional plateau and inclined segments.  
Note that this curve has a large 
$M=0$ plateau, which would require strong anisotropy and $\rho \ne 2$ to explain it within our model. 
Also note that transition from $M=0$ (Neel or dimer structure) 
to $M=1/3$ (UUD structure) is rather sharp in the experiment, while
there is a wide inclined segment between the $1/3$ plateau and the saturation.
From the experimental data we can roughly estimate the position of three most important points in the $M(h)$ curve:
the transition between $M=0$ and $M=1/3$ ($H_1 =1.79$ T),
end of the $M=1/3$ step ($H_2 = 2.6$ T), and saturation ($H_3 = H_{sat}=3.2$ T).
In Fig. \ref{f:hob4} the experimental magnetization curve $M(H/H_{sat})$ of HoB$_4$ (solid black curve)
is compared to the magnetization curves of Ising (dotted green line) and Heisenberg SSL
(red circles and dashed blue line). The latter ones has been calculated using our code
on the 6 x 2 lattice (the results were found to coincide with the ones obtained on the 6 x 6 and 12 x 12
lattices for the values of $\rho$ used).
Ising model can account for the sharp Neel/dimer--UUD transition, but it necessarily
predicts a sharp UUD--Spin Flip transition  as well, which contradicts the experiment.
Using the exact solution of the Ising SSL \cite{dublenych12prl109:167202}, we find that
the desired ratio $H_1/H_3$ is achieved for $\rho=0.6886$, or, alternatively, for $\rho=9.617$
(dotted green line in Fig. \ref{f:hob4}).
The $M=0$ plateau corresponds to Neel and dimer structure for these two cases respectively.
With the two parameters $\rho$, $\beta$ of the anisotropic Heisenberg SSL it is possible 
to reproduce the correct values of both ratios $H_1/H_3$ and $H_2/H_3$. It is achieved either 
for $\rho=0.4997$, $\beta=2.6359$, $H_{sat}=5.3641 \: J_1$ (red circles in Fig. \ref{f:hob4}), or
for $\rho=14.0$, $\beta=10.69$, $H_{sat}=21.453 \: J_1$
(dashed blue line in Fig. \ref{f:hob4}). The first set of values
provides better overall agreement with the experiment, as it gives a sharp first order Neel-UUD transition.
In fact, all main features of the experimental magnetization curve of HoB$_4$ are reproduced.
Note that both possible values of $\rho$ are quite far from the special point $\rho=2$.
It is important to stress that the values $\rho=0.4997$, $\beta=2.6359$ were obtained 
within the anisotropic SSL as effective parameters which give a best fit to the experimental curve 
$M(H/H_{sat})$ of HoB$_4$. In real HoB$_4$ the ratios $J_2/J_1$ and $B/J_1$ might have
slightly different values, as this material is likely to possess additional long-range exchange
and dipolar interactions which were ignored in our model.

It is interesting to compare the different physics of Ising and classical Heisenberg
SSL (the difference between these two models has also been discussed recently
in Ref. \onlinecite{huo12arxiv:1211.3872}).
The Ising SSL allows for collinear structures only, and its zero-temperature $M(h)$ curves
consist of vertical and horizontal segments only (plateaus and first order phase transition).
In contrast, magnetization curves for isotropic Heisenberg SSL are linear until saturation
thanks to noncollinear spin-flop and umbrella structures. The anisotropic Heisenberg SSL
combines features of both Ising and Heisenberg models. Its magnetization curves $M(h)$
can include both horizontal steps (corresponding to collinear spin structures) and
inclined regions (corresponding to noncollinear structures).
While the Ising approach might be sufficient for TmB$_4$, both types of regions
are clearly seen experimentally in HoB$_4$. 
The next logical step in the study of classical Heisenberg SSL
would be determining the phase diagram of the classical Heisenberg
SSL with easy-axis anisotropy. This problem is, however, beyond the scope of the present paper
and will be addressed in our future research.

\section{Conclusion}
We have found analytically the exact ground state of the classical Heisenberg SSL
in the external field $h$ and presented the phase diagram of this model in the $(\rho,h)$ plane.
The phase diagram includes the spin-flop phase for $\rho \le 1$, the umbrella phase
for $\rho > 1$, and the spin-flip phase for sufficiently large $h$. The phase transitions between
these three phases are of the second order. The zero-temperature
magnetization curves $M(h)$ are linear until saturation with no features.
For $\rho=2$ there is an additional degeneracy and
an infinite number of spin structures which share the energy and magnetization with the umbrella one.
In particular, UUD structure can be realized at a single point $(\rho,h)=(2,3)$. 

The effect of the easy-axis uniaxial anisotropy on $M(h)$ curves has been examined 
by numerical micromagnetic simulation. The anisotropy leads to the
onset of the $M=1/3$ UUD plateau at a certain finite value of the anisotropy
constant $\beta$ for $\rho \ne 2$ (and at $\beta=0$ for $\rho=2$).
Our results demonstrate the existence of both collinear (steps of the $M(h)$ curve)
and noncollinear (inclined parts of the $M(h)$ curve) spin structures for anisotropic
Heisenberg SSL. The results explain the magnetization curve of HoB$_4$, while
the $M(h)$ curves of TmB$_4$ are more Ising-like in nature, and cannot be
explained without introducing additional long-range interactions into the model.
\begin{acknowledgments}
The author thanks Prof. L.A. Pastur, Dr. V.V. Slavin,
and other colleagues for useful discussions.
\end{acknowledgments}
%


%

\end{document}